\documentclass[aps,pra,twocolumn,superscriptaddress,amsmath,amssymb,showpacs,notitlepage,longbibliography]{revtex4-1}
\pdfoutput=1
\usepackage{algorithm}
\usepackage{color}
\usepackage{algorithmic}
\usepackage{amsmath}
\usepackage{braket}
\usepackage{multirow}
\usepackage{amssymb}
\usepackage{amsthm}
\usepackage{bm}
\usepackage{thmtools}

\usepackage{array}
\usepackage{url}
\usepackage{hyperref}
\hypersetup{
    colorlinks=true,
    linkcolor=blue,
    filecolor=magenta,      
    urlcolor=cyan,
    citecolor=red
}
\usepackage{graphicx}
\usepackage{bibentry}
\usepackage{booktabs}
\usepackage{dcolumn}
\usepackage{bm}
\usepackage{blkarray}
\usepackage{xparse}
\usepackage{mathtools}
\usepackage{microtype}
\usepackage{enumerate}
\usepackage{soul}

\usepackage{makecell}

\providecommand{\U}[1]{\protect\rule{.1in}{.1in}}
\newtheorem*{theorem*}{Theorem}

\newtheorem{condition}{Condition}
\newtheorem{conjecture}{Conjecture}

\newcommand{\R}{\mathbb{R}}
\newcommand{\N}{\mathbb{N}}

\begin{document}

\title{Wigner non-negative states that verify the Wigner entropy conjecture}

\author{Qipeng Qian}
\affiliation{Program in Applied Mathematics, The University of Arizona, Tucson, Arizona 85721, USA}
\author{Christos N. Gagatsos}
\affiliation{Department of Electrical and Computer Engineering, The University of Arizona, Tucson, Arizona, 85721, USA}
\affiliation{Wyant College of Optical Sciences, The University of Arizona, Tucson, Arizona, 85721, USA}
\affiliation{Program in Applied Mathematics, The University of Arizona, Tucson, Arizona 85721, USA}

\begin{abstract}
We present further progress, in the form of analytical results, on the Wigner entropy conjecture set forth in [\href{https://link.aps.org/doi/10.1103/PhysRevA.104.042211}{Phys. Rev. A 104, 042211 (2021)}] and [\href{https://iopscience.iop.org/article/10.1088/1751-8121/aa852f/meta}{J. Phys. A: Math. Theor. 50 385301}]. Said conjecture asserts that the differential entropy defined for non-negative, yet physical, Wigner functions is minimized by pure Gaussian states while the minimum entropy is equal to $1+\ln\pi$. We prove this conjecture for the qubits formed by Fock states $|0\rangle$ and $|1\rangle$ that correspond to non-negative Wigner functions. In particular, we derive an explicit form of the Wigner entropy for those states lying on the boundary of the set of Wigner non-negative qubits. We then consider general mixed states and derive a sufficient condition for the conjecture's validity. Lastly, we elaborate on the states which are in accordance with our condition. 
\end{abstract}
\maketitle

\section{Introduction}\label{sec:intro}
Uncertainty relations are of fundamental interest in quantum information theory. They are closely related to the wave-particle duality in quantum mechanics and also illustrate one of the essential difference between quantum and classical mechanics. Furthermore, uncertainty relations directly put constraints on the precision of measurements and indicates inherent limitations in our understanding of quantum systems. 

The exploration of uncertainty relations traces back to the foundational Heisenberg uncertainty principle \cite{heisenberg1927anschaulichen}, in which the variance of the quadrature operators $\hat{q}$ and $\hat{p}$ is used as the quantifier of uncertainty. Later studies on uncertainty relations resulted in a natural generalization of classical information-related entropies to quantum systems (for an overview of entropic uncertainty relations see for example \cite{Coles2017}). In \cite{bialynicki1975uncertainty} an entropic uncertainty relation has been presented setting a lower bound on the summation of the Shannon entropy of the probability distribution function (PDF) of the position and the Shannon entropy of the PDF of the momentum of a quantum system. Said lower bound is stronger than the Heisenberg uncertainty relation. Furthermore, considering the subadditivity of Shannon’s differential entropy, one can expect that the entropy of the joint distribution of $q$ and $p$, i.e., of the Wigner function, will induce an even stronger bound, which would nevertheless imply the bound in \cite{bialynicki1975uncertainty}.

The Wigner entropy $S[W]$ is defined in \cite{VanHerstraeten2021quantum} as the differential Shannon entropy of the Wigner function $W(q,p)$ of the state with non-negative Wigner function (not necessarily corresponding to a classical state),
\begin{eqnarray} 
    \label{eq:defwigner}
    S[W]=-\int dqdp W(q,p)\ln W(q,p).
\end{eqnarray}
It possesses several properties \cite{VanHerstraeten2021quantum} such as additivity (for product states) and, unlike the Wehrl entropy \cite{Wehrl1979}, invariance under symplectic transformations. The Wigner entropy has been used in the analysis of noisy polarizers \cite{Aifer2023}, high energy physics \cite{Barata2023}, and non-equilibrium field theory \cite{Reiche2022}. In a broader view, phase space methods exploring the properties of quantum states are always of current interest, see for example \cite{Garttner2023,VanHerstraeten2023continuous,Salazar2023}. In this work, we focus on the following conjecture, presented in \cite{VanHerstraeten2021quantum,hertz2017,hertzthesis}, 
\begin{conjecture}
\label{Conj1}
For any Wigner non-negative state, 
\begin{eqnarray}
    \label{wignerlowbdd}
    S[W]\geq 1+\ln\pi,    
\end{eqnarray}
while the lower bound is attained by any pure Gaussian state.
\end{conjecture}

It is known that the marginals of the Wigner function coincide with probability densities of $q$ and $p$, denoted as $P_q=\int dp W(q,p)$ and $P_p=\int dq W(q,p)$ respectively. As discussed before, the Bialynicki-Birula-Mycielski inequality \cite{bialynicki1975uncertainty} and the subadditivity of Shannon’s differential entropy give, 
\begin{eqnarray}
    \label{Bialynicki}
    S[P_q]+S[P_p]&\geq& 1+\ln\pi,\\
 \label{eq:subadd}   S[P_q]+S[P_p]&\geq& S[W],
\end{eqnarray}
If Conjecture \ref{Conj1} is true, inequalities \eqref{Bialynicki} and \eqref{eq:subadd} can be written as,
\begin{eqnarray}
    S[P_q]+S[P_p]\geq S[W] \geq 1+\ln \pi,
\end{eqnarray}
i.e., we get a stronger entropic uncertainty relation for Wigner non-negative states.

In \cite{VanHerstraeten2021quantum}, Conjecture \ref{cond1} was proven analytically for passive states, i.e., states whose Fock basis representation has the form $\hat{\rho}_p = \sum_{n=0}^\infty q_n |n\rangle\langle n|$ under the condition  $\ q_{n+1} \leq q_n$, where $q_n$ are probabilities (non-negative real numbers satisfying $\sum_{n=0}^\infty q_n = 1$), and provided semi-numerical evidence for states that can be produced by mixing Wigner non-negative states in a balanced beam splitter. In this paper, we prove analytically Conjecture \ref{Conj1} for: (i) Qubits in the basis $\{|0\rangle, |1\rangle\}$, where $|0\rangle$ and $|1\rangle$ are Fock states, and (ii) for general mixed states which are Wigner non-negative and satisfy a sufficient condition. Throughout the paper we consider single-mode states.

It is worthwhile to mention that recent progress has been made in similar conjectures relating to the family of R\'enyi entropies for non-negative Wigner functions \cite{Kalka2023,Dias2023}.

This paper is organized as follows: In Section \ref{sec:qubit}, we analyze the conditions for any qubit in the basis $\{|0\rangle, |1\rangle\}$ to have a non-negative Wigner function. We then demonstrate that the Wigner entropy attains its minimum on the boundary of the Wigner non-negative set as defined by the derived condition. We explicitly derive the form of the Wigner entropy for these states and indeed verify that the lower bound of the Wigner entropy of such Wigner non-negative qubits matches the one of Conjecture \ref{Conj1}. In Section \ref{sec:generalwigner}, we consider the more general case of any mixed state and derive a sufficient condition such that Wigner non-negative states satisfy the lower bound stated in the Conjecture \ref{Conj1}. In Section \ref{sec:examples}, by showing a few concrete examples, we demonstrate that the set of states in accordance with our condition is non-empty and distinct from the set described in \cite{VanHerstraeten2021quantum}. Finally, in Section \ref{sec:conclusions} we summarize our results very briefly and we discuss future directions.

\section{Qubit states}\label{sec:qubit}

We denote the matrix representation of the density operator of a qubit formed by Fock states $|0\rangle$ and $|1\rangle$ in Bloch ball representation as,
\begin{eqnarray}
\label{eq:qubit}
\rho=\frac{1}{2}\left(I+r_1\sigma_x+r_2\sigma_y+r_3\sigma_z\right),
\end{eqnarray}
where $\{\sigma_x,\sigma_y,\sigma_z\}$ are the Pauli matrices,
\begin{eqnarray}
   \sigma_x = \begin{pmatrix}
       0 & 1 \\
1 & 0
   \end{pmatrix},\
   \sigma_{y}=\begin{pmatrix}
0 & -i \\
i & 0
\end{pmatrix},\
   \sigma_{z}=\begin{pmatrix}
1 & 0 \\
0 & -1
\end{pmatrix},
\end{eqnarray}
$I$ is the identity matrix, and $r_i\in[-1,1]$ for $i=1,2,3$ satisfy,
\begin{eqnarray}
\label{eq:BlochBall}    r_1^2+r_2^2+r_3^2\leq 1.
\end{eqnarray}
Following standard procedures (e.g. the compact formulation presented in \cite[Section 1.5 therein]{Ferraro2005}) we can find the Wigner function corresponding to Eq. \eqref{eq:qubit} to be,
\begin{eqnarray}
\nonumber W(q,p)&=&\frac{1}{\pi}e^{-q^2-p^2}\big[ (q^2+p^2)(1-r_3)\\
 \label{eq:WNN-qubit}   &&+\sqrt{2}r_1q+\sqrt{2}r_2p+r_3 \big].
\end{eqnarray} 
Since Eq. \eqref{eq:qubit} under condition \eqref{eq:BlochBall} represents a physical state, the Wigner function of Eq. \eqref{eq:WNN-qubit} is naturally physical under the same condition. However, we need to identify a condition on $r_i,\ i=1,2,3$, such that Eq. \eqref{eq:WNN-qubit} also represents a non-negative Wigner function. To this end, we require $W(q,p)\geq 0$ and we derive the following condition,
\begin{eqnarray}
    \label{eq:nonnega-WNN-qubit}
    2(r_1^2+r_2^2)+(1-2r_3)^2\leq1.
\end{eqnarray}
Under the conditions of Eqs. \eqref{eq:BlochBall} and \eqref{eq:nonnega-WNN-qubit}, we are able to analyze whether the Conjecture \ref{Conj1} is true for the qubit case. However, some observations leading to simplifications are due.

First, the Wigner entropy is invariant under symplectic transformations. Therefore, we can set $r_2=0$ in Eq. \eqref{eq:WNN-qubit} since arbitrary values of $r_2$ correspond to optical phase shifts, i.e., a Gaussian unitary transformation \footnote{This can be proven as follows: Use spherical coordinates in Eq. \eqref{eq:qubit}, i.e., for $|r|\in [0,1]$, $\theta \in [0,\pi]$, $\phi \in [0,2\pi)$, we write $r_1=|r| \sin \theta \cos \phi$, $r_2=|r| \sin \theta \sin \phi$, and $r_3=|r| \cos \theta$. Then, applying the (Gaussian unitary) phase shift operator $e^{-i \phi \hat{n}}$ on the state of Eq. \eqref{eq:qubit}, i.e., calculating $e^{-i \phi \hat{n}} \rho e^{i \phi \hat{n}}$ and using the fact that the Pauli operators are represented as matrices on the $\{|0\rangle,|1\rangle\}$ (Fock states) basis, we find that the resulting state to be $|r|(I+\sin\theta \sigma_1+\cos\theta \sigma_3)/2$ or equivalently in Cartesian coordinates $(I+ r_1 \sigma_1+r_3 \sigma_3)/2$.} (optical phase shifting corresponds to a symplectic transformation on phase space). Therefore, the Wigner function and the conditions we consider, respectively become, 
\begin{eqnarray}
\nonumber   W_{13}(q,p) &=&\frac{1}{\pi}e^{-q^2-p^2}[ (q^2+p^2)(1-r_3)\\
  \label{eq:WNN-qubit2}   &&+\sqrt{2}r_1 q+r_3 ],
\end{eqnarray}
\begin{eqnarray}
    \label{eq:BlochDisk}
    r_1^2+r_3^2\leq1,
\end{eqnarray}
\begin{eqnarray}
    \label{eq:nonnega-WNN-qubit2}
    2 r_1^2+(1-2r_3)^2\leq1.
\end{eqnarray}

\begin{figure*}
  \includegraphics[width=0.7\textwidth]{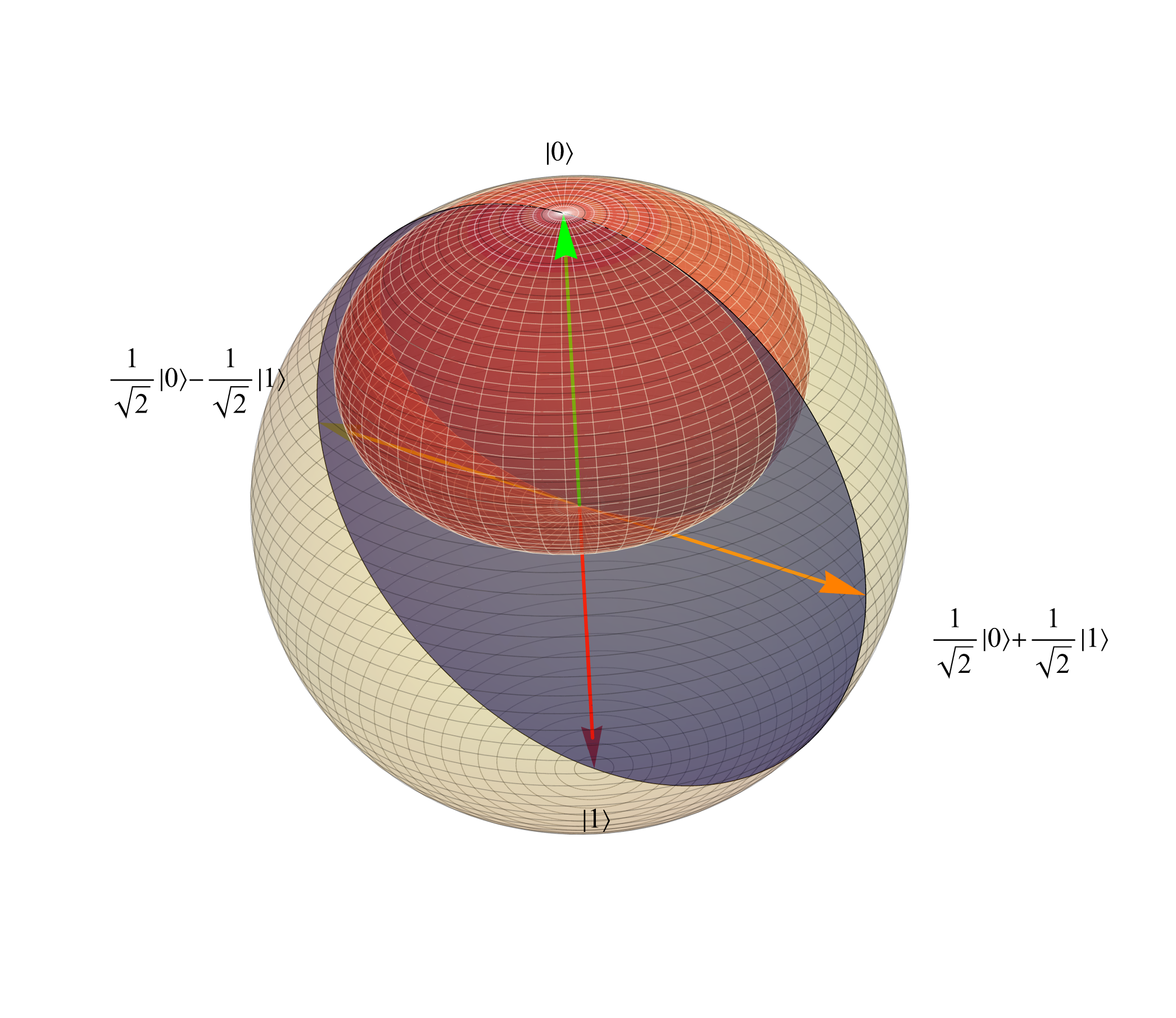}
  \caption{The Bloch ball (yellow with black mesh lines), as defined by the $|0\rangle$ and $|1\rangle$ Fock vectors, contains all physical states, even those corresponding to partly negative Wigner functions. The ellipsoid (red with white mesh lines) contains physical states whose Wigner function is non-negative. States that belong to the same latitude are equivalent up to an optical phase. Therefore, it suffices to look only at states contained in the intersection of the disk (shown in blue) and the ellipsoid (the region defined by Eq. \eqref{eq:nonnega-WNN-qubit2}), i.e., physical states with non-negative Wigner functions. In fact, it suffices to consider only states contained in said elliptical region only on the left of (or on the right of) the line depicting the $|0\rangle$ vector. This is because symmetric states with respect to the upward vector are equivalent up to an optical phase shifting.}
  \label{fig:Bloch}
\end{figure*}

The Bloch ball with the Wigner non-negative regions of our system is depicted in Fig. \ref{fig:Bloch}.

Second, for any fixed $q,p,r_3$, the second derivative on $-W_{13}(q,p)\ln W_{13}(q,p)$ with respect to $r_1$ gives,
\begin{eqnarray} 
 \nonumber   \frac{\partial^2}{\partial r_1^2}\left[-W_{13}(q,p)\ln(W_{13}(q,p))\right]
    \\
\label{eq:convexity-Wr}    =-\frac{2q^2 e^{-2q^2-2p^2}}{\pi^2W_{13}(q,p)}
    \leq0,
\end{eqnarray}
implying that the Wigner entropy is concave with respect to $r_1$ due to the linearity of integration in Eq. \eqref{eq:defwigner}. Thus, the minimum of the Wigner entropy can only exist at some point along the boundary defined by the condition \eqref{eq:nonnega-WNN-qubit2}, i.e.,
\begin{eqnarray}
    \label{eq:bdd-qubit}
    2r_1^2+(1-2r_3)^2=1.
\end{eqnarray}

\begin{figure*}
  \includegraphics[width=0.7\textwidth]{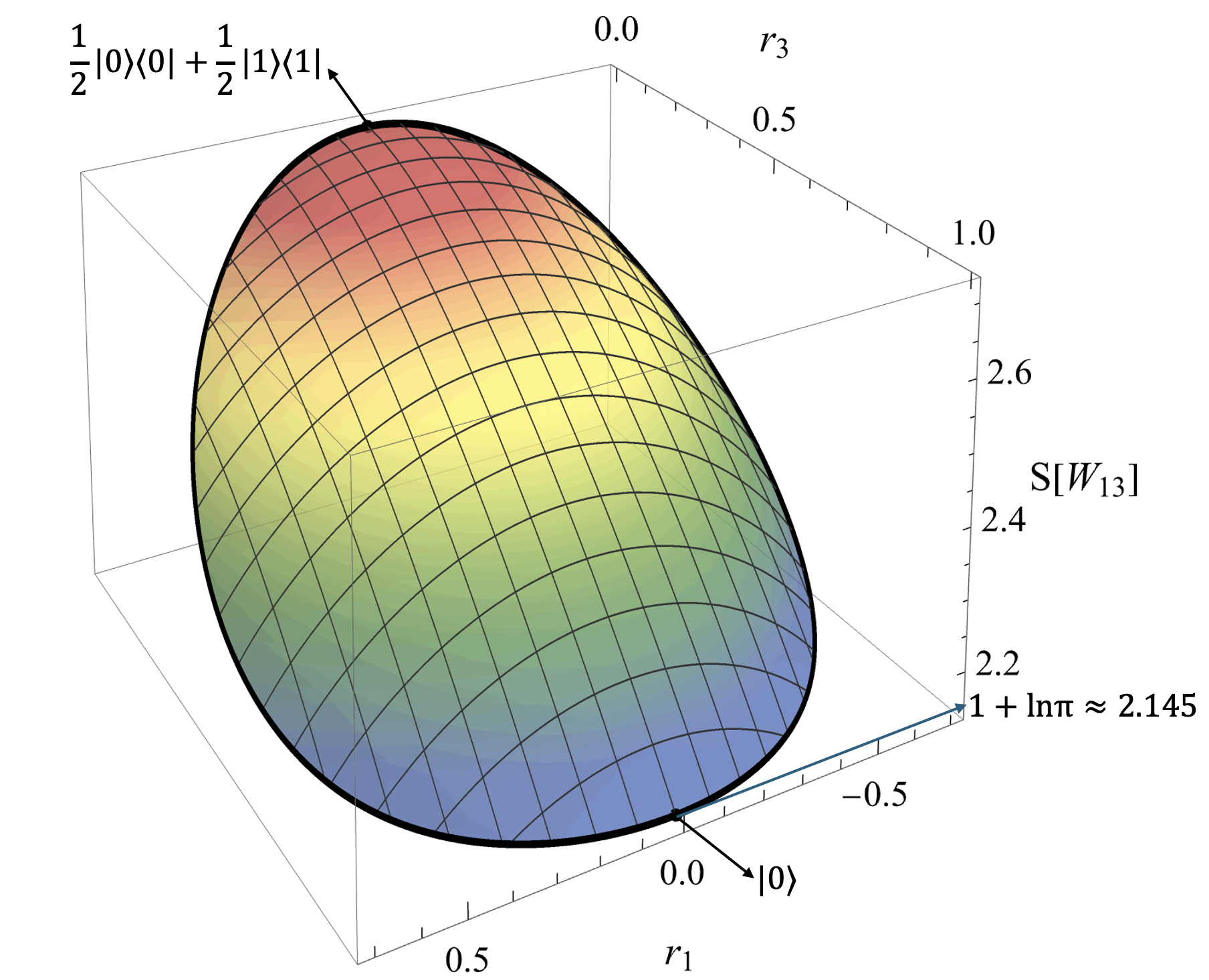}
  \caption{The Wigner entropy $S[W_{13}]$ numerically evaluated for $20,000$ unique choices of $(r_1,r_3)$, compatible with Eqs. \eqref{eq:BlochDisk} and \eqref{eq:nonnega-WNN-qubit2}. The border of the surface is the Wigner entropy $S[W_3^{\pm}]$ (its analytical expression is given in Eq. \eqref{eq:S_3}), while the upper point corresponds to the maximally mixed state $\frac{1}{2}|0\rangle \langle 0| + \frac{1}{2}|1\rangle \langle 1|$ and lower point corresponds to $|0\rangle$. The concavity with respect to $r_1$ and the minimization of the Wigner entropy are proven analytically in the main text.}
  \label{fig:Sw13}
\end{figure*}

Finally, using Eq. \eqref{eq:bdd-qubit} we simplify further Eq. \eqref{eq:WNN-qubit2},
\begin{eqnarray}
\nonumber   W_{3}^{\pm}(q,p) &=&\frac{1}{\pi}e^{-q^2-p^2}[ (q^2+p^2)(1-r_3)\\
  \label{eq:W3pm}   &&\pm 2 q \sqrt{r_3 (1-r_3)}+r_3 ],
\end{eqnarray}
where the $\pm$ corresponds to the two possible solutions of Eq. \eqref{eq:bdd-qubit} with respect to $r_1$. Both Wigner functions of Eq. \eqref{eq:W3pm} are equivalent up to an optical phase. Therefore, they correspond to the same Wigner entropy. We choose to work with $W_{3}^+(q,p)$,
\begin{eqnarray}
  \label{eq:W3}  W_{3}(q,p) \equiv W_{3}^{+} (q,p).
\end{eqnarray}

The explicit form of the Wigner entropy on the boundary defined by Eq. \eqref{eq:bdd-qubit} (see Appendix \ref{appdxA}) is, 
\begin{eqnarray} 
 \nonumber  S_b\equiv S [W_{3}^{\pm}]&=&e^{-\frac{r_3}{1-r_3}}(1-r_3)+r_3+\ln\frac{\pi}{r_3}\\
     \label{eq:S_3}   &&+\text{Ei}\left(-\frac{r_3}{1-r_3}\right),
\end{eqnarray}
where $\text{Ei}(x)=\int_{-\infty}^x dt \frac{e^t}{t}$ is the exponential integral function and the subscript $b$ denotes that we work on the boundary of the Wigner non-negative qubits. In Fig. \ref{fig:Sw13}, we plot the Wigner entropy $S[W_{13}]$. We find that the entropy $S_b$ is minimized at $r_3=1$ and its minimum value is $1+\ln\pi$, which is consistent with the Conjecture \ref{Conj1}. For $r_3=1$ and $r_2=0$, per Eq. \eqref{eq:BlochBall} we get $r_1=0$, which means that the state minimizing the Wigner entropy is $|0\rangle$, which is the only pure Gaussian state in the Bloch ball. The details on the minimization are provided in Appendix \ref{appdxA}.

\section{A subset of Wigner non-negative states}\label{sec:generalwigner}
In this section, for a restricted set of physical and Wigner non-negative (in general mixed) states $\hat{\rho}$, we find that the Wigner entropy is in accordance with Conjecture \ref{Conj1}.

Let $k\in\R$ and $k\geq1$. For any non-negative Wigner function $W(q,p)$, we define the functional, 
\begin{eqnarray} 
    \label{eq:mu_k}
    \mu_k[W]=k\pi^{k-1}\int dqdpW^k(q,p). 
\end{eqnarray}

\begin{conjecture}
\label{Conj2}
\begin{eqnarray} 
    \label{eq:Conj2}
    \frac{\partial \mu_{k}[W]}{\partial k} \Big|_{k\to 1} \leq 0.
\end{eqnarray}
\end{conjecture}
The left hand side of Eq. \eqref{eq:Conj2} is equal to $1+\ln\pi-S[W]$. Therefore, Conjecture \ref{Conj2} is equivalent to Conjecture \ref{Conj1}. Additionally, for $k=1$ we get the normalization property of Wigner functions, 
\begin{eqnarray}
    \label{eq:mu_1}
    \mu_1[W]=\int dpdqW(q,p)=1,
\end{eqnarray}
Therefore, Conjecture \ref{Conj2} is true if the following sufficient condition holds, 
\begin{eqnarray}
    \label{eq:condition-mu_k}
    \mu_k[W]\leq 1
\end{eqnarray}
for $k\in[1,2]$. Furthermore, utilizing the fact that 
\begin{eqnarray} 
    \int dqdpW_0^k(q,p)=\frac{1}{k\pi^{k-1}},
\end{eqnarray}
where $W_0(q,p)$ denotes the Wigner function of $|0\rangle$, we can rewrite Eq. \eqref{eq:condition-mu_k} as,
\begin{eqnarray} 
    \label{eq:condition-nu_k}
    \nu_k[W]\leq \nu_k[W_0],
\end{eqnarray}
where,
\begin{eqnarray} 
    \label{eq:nu_k}
    \nu_k[W]=\int dqdpW^k(q,p)
\end{eqnarray}
for $k\in[1,2]$.

We now derive a sufficient condition such that Eq. \eqref{eq:condition-nu_k} is true. For any (generally mixed) state $\hat{\rho}$, its Wigner function has the form 
\begin{eqnarray} 
    \label{eq:general-W}
    W_{\hat{\rho}}(q,p)=W_0(q,p)P(q,p),
\end{eqnarray}
where $P(q,p)=\sum_{a,b=0}^{\infty}c_{ab}q^ap^b$ is a polynomial in $q,p$. To prove Eq. \eqref{eq:general-W}, one can start by writing $\hat\rho$ on the Fock basis and then calculate the generic Wigner function using well-known methods (see for example \cite{Ferraro2005}). Since Wigner functions are real-valued, $c_{ab}\in\R$ for any $(a,b)\in\mathbb{N}^2$. To make $W(q,p)$ normalized, we can rewrite $W(q,p)$ as 
\begin{eqnarray} 
    \label{eq:general-normal-W}
    W(q,p)=W_0(q,p) \sum_{a,b=0}^\infty\frac{\pi\Tilde{c}_{ab}}{\Gamma\left(\frac{1+a}{2}\right)\Gamma\left(\frac{1+b}{2}\right)}q^ap^b,
\end{eqnarray} 
where $\Gamma(x)=\int_0^{\infty}s^{x-1}e^{-s}ds$ and 
\begin{eqnarray}
        \sum_{a,b=0}^{\infty}\tilde{c}_{ab}&=&1.
\end{eqnarray}
We define $F_{ab}(q,p)$ as,
\begin{eqnarray} 
    \label{eq:Fab}
    F_{ab}(q,p)
    =W_0(q,p)\frac{\pi}{\Gamma\left(\frac{1 + a}{2}\right) \Gamma\left(\frac{1 + b}{2}\right)}q^ap^b.
\end{eqnarray}
In Appendix \ref{appdxB} we prove that,
\begin{eqnarray} 
    \label{even}
    \|F\|_k^k\leq \nu_k[W_0],
\end{eqnarray}
where $\|F\|_k=\left(\int dqdp |F_{ab}(q,p)|^k\right)^{\frac{1}{k}}$ for any $(a,b)\in\N^2$ and $k\in[1,2]$.

Let us impose the following condition,
\begin{condition}
\label{cond1}
$$\sum_{a,b=0}^{\infty}|\Tilde{c}_{ab}|=1.$$
\end{condition} 
\noindent For a non-negative Wigner function $W(q,p)$ satisfying Condition \ref{cond1}, we utilize Eq. \eqref{even} and the triangle inequality to get,
\begin{eqnarray} 
    \nu_k[W] \equiv \|W\|_k^k
    &\leq&\left[\sum_{a,b=0}^{\infty}|\Tilde{c}_{ab}|\|F_{ab}(q,p)\|_k\right]^k  \\
    &\leq&\left[\sum_{a,b=0}^{\infty}|\Tilde{c}_{ab}|(\nu_k[W_0])^{\frac{1}{k}}\right]^k\\
    &=&\left[\left(\nu_k[W_0]\right)^{\frac{1}{k}}\right]^k=\nu_k[W_0].
\end{eqnarray}
Therefore, Condition \ref{cond1} is a sufficient condition for Conjecture \ref{Conj2} and thus Conjecture \ref{Conj1} to hold. 

\section{Examples}\label{sec:examples}
We provide a few examples demonstrating that the set of states satisfying our conditions is non-empty and distinct from the set of passive states explored in \cite{VanHerstraeten2021quantum}. In particular, we give three examples to show how the set of states explored in Section \ref{sec:generalwigner} intersects with both the sets of passive states and of Fock-diagonal states.

\emph{Example 1:} Consider states of the form $p_0|0\rangle\langle0|+p_1|1\rangle\langle1|$ with $0 \leq p_i\leq 1,\ i=0,1$, and $\sum_{i=0}^1 p_i=1$. All states of this form which additionally satisfy $p_1\leq\frac{1}{2}$ (e.g. passive states) satisfy Condition \ref{cond1}. 

\emph{Example 2:} Consider states of the form $p_0|0\rangle\langle0|+p_1|1\rangle\langle1|+p_2|2\rangle\langle2|$ with $0 \leq p_i\leq 1,\ i=0,1,2$, and $\sum_{i=0}^2 p_i=1$. All such states with
\begin{eqnarray}
  p_1&\leq&\frac{1}{2},\\
  \quad p_1-2p_2&\geq&0
\end{eqnarray}
satisfy Condition \ref{cond1}. This set of states does not necessarily include passive states. For example, for
\begin{eqnarray}
   p_0=p_2&=&\frac{1}{4},\\
   p_1&=&\frac{1}{2}
\end{eqnarray}
the state is not passive but still satisfies Condition \ref{cond1} and thus satisfies the Conjecture \ref{Conj1}.

\emph{Example 3:} Consider the state of $p_0|0\rangle\langle0|+p_1|1\rangle\langle1|+p_2|2\rangle\langle2|+p_3|3\rangle\langle3|$ with $0 \leq p_i\leq 1,\ i=0,1,2,3$, and $\sum_{i=0}^3 p_i=1$. All such states with,
\begin{eqnarray}
    p_1-2p_2+3p_3&\geq&0,\\
    p_2-3p_3&\geq&0,\\
    p_0+p_2&\geq&\frac{1}{2}
\end{eqnarray}
satisfy Condition \ref{cond1}. We note that when $p_0=p_1=p_2=p_3=\frac{1}{4}$, the state is passive but does not satisfy Condition \ref{cond1}. 

From the examples above, we observe that our set only intersects with the set of passive states but does not contain it. Moreover, it directly follows that if a state is Fock-diagonal and Wigner non-negative, the state is not necessarily in compliance with Condition \ref{cond1}. In Appendix \ref{appdxC}, we delve deeper into the relationship between our set and the set of Fock-diagonal states. We find that our Condition \ref{cond1} does not imply that a state $\hat{\rho}$ is Fock-diagonal but it does imply it when $\rho_{n,m}=0$ if $|n-m|$ is odd, where $\rho_{n,m}$ is the element of the matrix representation of $\hat{\rho}$ on the Fock basis.

\section{Conclusions}\label{sec:conclusions}
In this paper, we proved that the newly introduced Conjecture \ref{Conj1} \cite{VanHerstraeten2021quantum,hertz2017,hertzthesis} holds true for two cases: for (generally mixed) qubits formed by Fock states $|0\rangle$ and $|1\rangle$ and for states that satisfy Condition \ref{cond1}. Therefore, for Wigner non-negative states, we presented progress towards a stronger position-momentum uncertainty relation compared to the one derived in the seminal work \cite{bialynicki1975uncertainty} and expanded the results of \cite{VanHerstraeten2021quantum}. Moreover, the entropic uncertainty relation considered in this work, subsumes \cite{VanHerstraeten2021quantum} the Wehrl entropy inequality for the always non-negative $Q$ functions. We note that the Wehrl entropy is minimized for coherent states while it is not in general invariant under symplectic transformations, e.g., the Wehrl entropy of a coherent state and a squeezed state are not in general equal.

\begin{figure}
  \centering
  \includegraphics[width=0.4\textwidth]{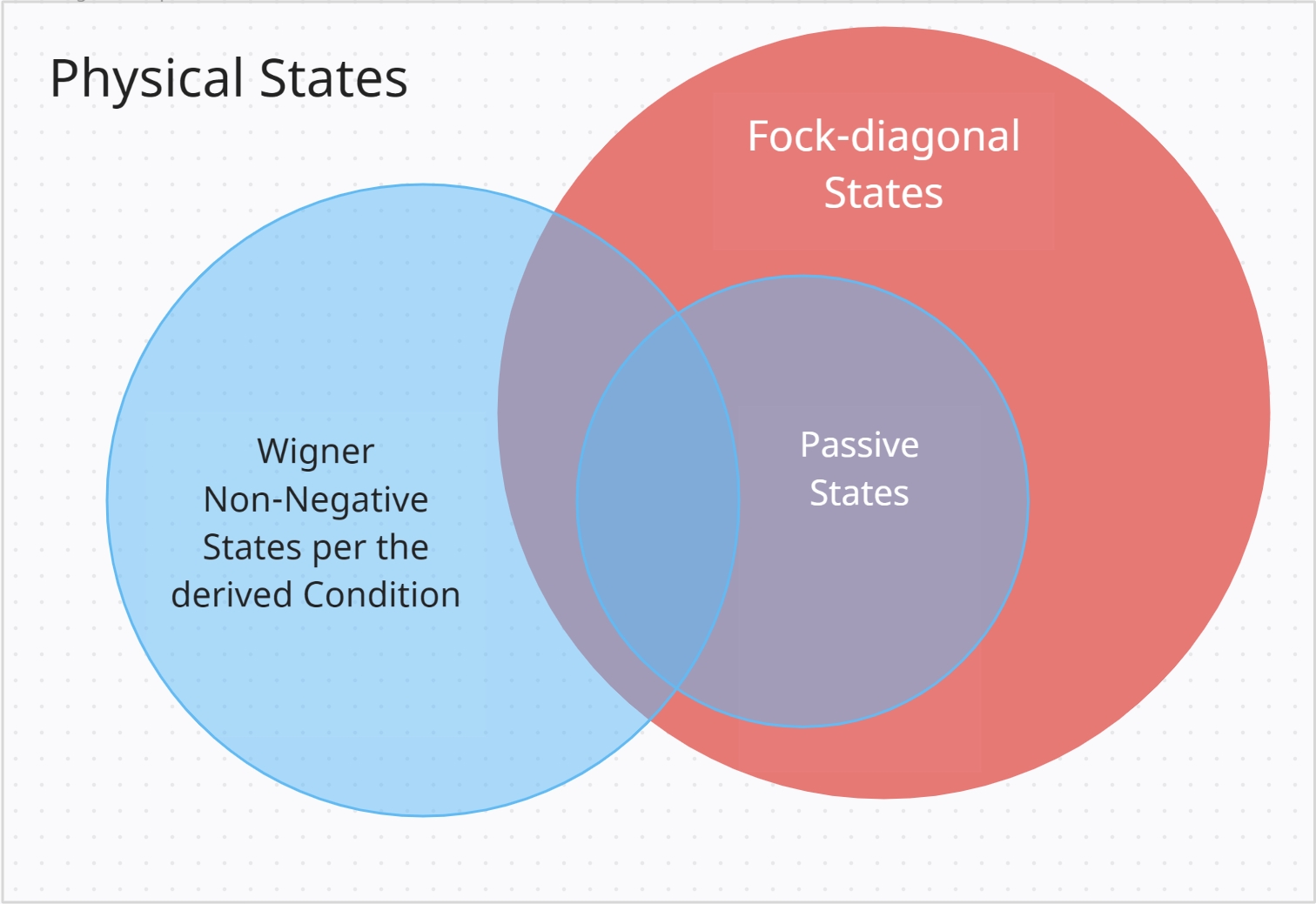}
  \caption{Graphical explanation of the relation between the states satisfying Condition \ref{cond1}, passive and (more generally) Fock-diagonal states.}
  \label{fig:set-relation}
\end{figure}

The relationship between the set defined by Condition \ref{cond1} and the sets of passive states and Fock-diagonal states is depicted in Fig. \ref{fig:set-relation}. 
The difficulty of proving Conjecture \ref{Conj1} for all $W(q,p)\geq 0$ lies in lacking a computationally-useful criterion for Wigner non-negativity which also excludes non-physical states: The condition $W(q,p)\geq 0$ merely imposes non-negativity on the the function, while one would need to take into account the condition,
\begin{eqnarray}
 \label{eq:CondPhys}   \int dq dp W(q,p) W_{|\psi\rangle}(q,p) \geq 0
\end{eqnarray}
\emph{for all pure states} $|\psi\rangle$, as well to ensure that $W(q,p)$ corresponds to a physical state. One way forward could be to consider a set of functions $\tilde{W}(q,p)$ that includes all physical Wigner functions, plus a subset of functions that are non-negative but do not correspond to physical states. For example, this can be done by considering only a (convenient) subset of pure states satisfying Eq. \eqref{eq:CondPhys}. We note that an approach leading to the conclusion that the Wigner non-negative state minimizing the Wigner entropy is a pure state would prove Conjecture \ref{Conj1} in general. This would be an immediate consequence of Hudson's theorem stating that any pure state with a non-negative Wigner function is necessarily a Gaussian state \cite{Hudson1974}. 

Lastly, we envision future works elaborating on: Conjecture \ref{Conj1} for states defined across multiple modes, entropy power inequalities \cite{Hertz_2019}, or even on the properties for the complex-valued Wigner entropy, i.e., for partly negative Wigner functions, in the direction of \cite{hertzthesis,pizzimenti2024exploring,Cerf2024}.

\begin{acknowledgements}
The authors are thankful to Zacharie Van Herstraeten, Michael G. Jabbour, Anaelle Hertz, and Nicolas Cerf for numerous fruitful discussions. C.N.G. and Q.Q. acknowledge financial support from the National Science Foundation, FET, Award No. 2122337.

\end{acknowledgements}

\newpage
\onecolumngrid
\appendix
\section{Minimization of Eq. \eqref{eq:S_3}}\label{appdxA}
\renewcommand{\thesubsection}{\arabic{subsection}}
\def\theequation{A\arabic{equation}}
\setcounter{equation}{0}
\renewcommand{\thefigure}{A\arabic{figure}}    
\setcounter{figure}{0}
In this Appendix, we prove that the Wigner entropy $S_b$ of Eq. \eqref{eq:S_3} attains its minimum at $r_3=1$ and the corresponding value is $1+\ln\pi$. 

Under the change of variables, 
 \begin{eqnarray}
    q'=\sqrt{1-r_3}q\ \text{and}\ p'=\sqrt{1-r_3}p
\end{eqnarray}
and using the condition of Eq. \eqref{eq:bdd-qubit}, Eq. \eqref{eq:W3} and its corresponding Wigner entropy become, 
\begin{eqnarray} 
 \label{eqA:W3}   W'_{3}(q',p')&=&\frac{1}{\pi}e^{-\frac{q'^2+p'^2}{1-r_3}}[p'^2+(q'+\sqrt{r_3})^2],\\
 \label{eqA:Sbprime}   S_b\equiv S[W'_3]&=&-\frac{1}{1-r_3}\int dq'dp'W'_3\ln(W'_3).
\end{eqnarray}
With further change of variables,  
\begin{eqnarray}
    q'=R\sin\theta\text{ and }p'=R\cos\theta,
\end{eqnarray}
satisfying $dq'dp'=RdRd\theta$, Eq. \eqref{eqA:W3} becomes,  \begin{eqnarray}\label{W3R}
    W''_{3}(R,\theta)=\frac{1}{\pi}e^{-\frac{R^2}{1-r_3}}[R^2+2\sqrt{r_3}R\sin\theta+r_3].
\end{eqnarray}
and the Wigner entropy $S_b$ of Eq. \eqref{eqA:Sbprime} in the new variables can be computed as, 
\begin{eqnarray}
    S_b \equiv S[W''_3]
    &=&-\frac{1}{1-r_3}\int_0^{\infty}\int_0^{2\pi}dRd\theta R\frac{1}{\pi}e^{-\frac{R^2}{1-r_3}}(R^2+2\sqrt{r_3}R\sin\theta+r_3)\ln(W_3)\\
    &=&-\frac{1}{1-r_3}\int_0^{\infty}\int_0^{2\pi}dRd\theta R\frac{1}{\pi}e^{-\frac{R^2}{1-r_3}}(R^2+2\sqrt{r_3}R\sin\theta+r_3)\\\nonumber
    &&\times\left[(-\ln\pi-\frac{R^2}{1-r_3})+\ln(R^2+r_3+2\sqrt{r_3}R\sin\theta)\right]\\
    &=&-\frac{1}{1-r_3}\int_0^{\infty}\int_0^{2\pi}dRd\theta R\frac{1}{\pi}e^{-\frac{R^2}{1-r_3}}\{(R^2+r_3)(-\ln\pi-\frac{R^2}{1-r_3})\\\nonumber
    &&+2\sqrt{r_3}R\sin\theta(-\ln\pi-\frac{R^2}{1-r_3})+(R^2+r_3)\ln(R^2+r_3+2\sqrt{r_3}R\sin\theta)\\
    &&+2\sqrt{r_3}R\sin\theta\ln(R^2+r_3+2\sqrt{r_3}R\sin\theta)\}.
\label{eqA:S3R}
\end{eqnarray}
We have the following useful relations pertaining to the previous integrals, 
\begin{eqnarray}
&\int_0^{2\pi}d\theta&\sin\theta=0,\\
&\int_0^{2\pi}d\theta&\ln(R^2+r_3+2\sqrt{r_3}R\sin\theta)=2\pi\left(\ln\frac{R^2+r_3}{2}+\ln\frac{R^2+r_3+\sqrt{(R^2-r_3)^2}}{R^2+r_3}\right),\\
&\int_0^{2\pi}d\theta&\sin\theta\ln(R^2+r_3+2\sqrt{r_3}R\sin\theta)=2\pi\frac{R^2+r_3-\sqrt{(R^2-r_3)^2}}{2\sqrt{r_3}R},
\end{eqnarray}
which when used in Eq. \eqref{eqA:S3R} we get,
\begin{eqnarray}
    S_b
    =2e^{-\frac{r_3}{1-r_3}}(1-r_3)+r_3-\frac{2U}{1-r_3},
    \label{S3}
\end{eqnarray}
where 
\begin{eqnarray}
    \frac{2U}{1-r_3}
    =-e^{-\frac{r_3}{1-r_3}}(r_3-1)-\text{Ei}(\frac{r_3}{-1+r_3})-\ln\frac{\pi}{r_3}.
\end{eqnarray}
Therefore, 
\begin{eqnarray}
S_b=e^{-\frac{r_3}{1-r_3}}(1-r_3)+r_3+\ln\frac{\pi}{r_3}+\text{Ei}(-\frac{r_3}{1-r_3}).
\end{eqnarray}
Then, we calculate the derivative 
\begin{eqnarray}
    \frac{d}{dr_3}S_b
    &=&e^{-\frac{r_3}{1-r_3}}\frac{r_3-2}{1-r_3}+1-\frac{1}{r_3}+\frac{e^{-\frac{r_3}{1-r_3}}}{r_3(1-r_3)}\\
    &=&e^{-\frac{r_3}{1-r_3}}\frac{r_3^2-2r_3+1}{r_3(1-r_3)}-\frac{1-r_3}{r_3}\\
    &=&e^{-\frac{r_3}{1-r_3}}\frac{1-r_3}{r_3}-\frac{1-r_3}{r_3}\\
    &=&\frac{1-r_3}{r_3}(e^{-\frac{r_3}{1-r_3}}-1)\\
    &\leq&0. 
\end{eqnarray}
It is clear that $\frac{d}{dr_3}S_b=0$ is only possible when $r_3=0,1$. By L'Hôpital's rule, we have 
\begin{eqnarray}
    &\lim_{r_3\rightarrow0}&\frac{(1-r_3)(e^{-\frac{r_3}{1-r_3}}-1)}{r_3}=\frac{1+\frac{e^{-\frac{r_3}{1-r_3}}(r_3-2)}{1-r_3}}{1}=-1,\\
    &\lim_{r_3\rightarrow1}&\frac{(1-r_3)(e^{-\frac{r_3}{1-r_3}}-1)}{r_3}=\frac{0(0-1)}{1}=0.
\end{eqnarray}
Therefore, we conclude that $S_b$ obtain its minimum value $1+\ln\pi$ at $r_3=1$.

\section{Proof of Eq. \eqref{even}}\label{appdxB}
\renewcommand{\thesubsection}{\arabic{subsection}}
\def\theequation{B\arabic{equation}}
\setcounter{equation}{0}
\renewcommand{\thefigure}{B\arabic{figure}}    
\setcounter{figure}{0}
 First, by plugging in $F_{ab}(q,p)$ to Eq. \eqref{even}, we get 
\begin{eqnarray}
    &\ln\left( \pi^{k-1}k^{-\frac{a+b}{2}k}\frac{\Gamma\left(\frac{1+ak}{2}\right)\Gamma\left(\frac{1+bk}{2}\right)}{\left[\Gamma\left(\frac{1+a}{2}\right)\Gamma\left(\frac{1+b}{2}\right)\right]^k} \right)\leq0,
\end{eqnarray}
which is equivalent to 
\begin{eqnarray}
\nonumber    &&(k-1)\ln\pi-\frac{a+b}{2}k\ln k+\ln\left(\Gamma\left(\frac{1+ak}{2}\right)\right)+\ln\left(\Gamma\left(\frac{1+bk}{2}\right)\right)-k\ln\left(\Gamma\left(\frac{1+a}{2}\right)\right)-\\&& k\ln\left(\Gamma\left(\frac{1+b}{2}\right)\right)\leq0.
\end{eqnarray}
Denote by $f(a,b,k)$ the left hand side of the above inequality and use the result from \cite{gordon1994stochastic},
\begin{eqnarray}
    \psi(x)\leq\ln x-\frac{1}{2x},
\end{eqnarray}
where $\psi(x)=\frac{d }{dz}\left[\ln\left(\Gamma\left(x\right) \right)\right]$, we get  
\begin{eqnarray}
    \frac{\partial f(a,b,k)}{\partial k}
    &=&\ln\pi-\frac{a+b}{2}(1+\ln k)+\frac{a}{2}\psi\left(\frac{1+ak}{2}\right)+\frac{b}{2}\psi\left(\frac{1+bk}{2}\right)\\\nonumber
    &&-\ln\left(\Gamma\left(\frac{1+a}{2}\right)\right)-\ln\left(\Gamma\left(\frac{1+b}{2}\right)\right)\\
    &\leq&\ln\pi-\frac{a+b}{2}(1+\ln k)+\frac{a}{2}\ln\frac{1+ak}{2}-\frac{a}{2(1+ak)}\\\nonumber
    &&+\frac{b}{2}\ln\frac{1+bk}{2}-\frac{b}{2(1+bk)}-\ln\left(\Gamma\left(\frac{1+a}{2}\right)\right)-\ln\left(\Gamma\left(\frac{1+b}{2}\right)\right)\\
    &=:&g(a,b,k).
\end{eqnarray}
We can then calculate 
\begin{eqnarray}
    \frac{\partial g(a,b,k)}{\partial k}
    &=&-\frac{a+b}{2k}+\frac{a^2}{2(1+ak)}+\frac{a^2}{2(1+ak)^2}+\frac{b^2}{2(1+bk)}+\frac{b^2}{2(1+bk)^2}\\
    &=&-\frac{a}{2k(1+ak)^2}-\frac{b}{2k(1+bk)^2}<0,
\end{eqnarray}
where we always work with $k\in[1,2]$.

Denote $h(a,b)$ as 
\begin{eqnarray}
    h(a,b)
    &=&g(a,b,1)\\
    &=&\ln\pi-\frac{a+b}{2}+\frac{a}{2}\ln\frac{1+a}{2}-\frac{a}{2(1+a)}+\frac{b}{2}\ln\frac{1+b}{2}\\\nonumber
    &&-\frac{b}{2(1+b)}-\ln\left(\Gamma\left(\frac{1+a}{2}\right)\right)-\ln\left(\Gamma\left(\frac{1+b}{2}\right)\right).
\end{eqnarray}
When $a>0$, we apply results from \cite{gordon1994stochastic},
\begin{eqnarray}
    \psi(x)>\ln x-\frac{1}{2x}-\frac{1}{12x^2},
\end{eqnarray}
to get 
\begin{eqnarray}
    \frac{\partial h(a,b)}{\partial a}
    &=&-\frac{1}{2}+\frac{1}{2}\ln\frac{1+a}{2}+\frac{a}{2(1+a)}-\frac{1}{2(1+a)^2}-\frac{1}{2}\psi\left(\frac{1+a}{2}\right)\\
    &=&\frac{1}{2}\left( -1+\ln\frac{1+a}{2}+\frac{a}{1+a}-\frac{1}{(1+a)^2}-\psi\left(\frac{1+a}{2}\right)\right)\\
    &<&\frac{1}{2}\left( -1+\ln\frac{1+a}{2}+\frac{a}{1+a}-\frac{1}{(1+a)^2}-\ln\frac{1+a}{2}+\frac{1}{1+a}+\frac{1}{3(1+a)^2}\right)\\
    &=&-\frac{1}{3(1+a)^2}\\
    &<&0.
\end{eqnarray}
Similar for $b$, we have $\frac{\partial h(a,b)}{\partial b}<0$. 

Now, set $a$ or $b$ equal to $0$, we also have 
\begin{eqnarray}
    \frac{\partial h(a,b)}{\partial a}\Bigg|_{a=0}&<&0\\
    \frac{\partial h(a,b)}{\partial b}\Bigg|_{b=0}&<&0.
\end{eqnarray}
Thus, for any $a,b\in\N$, we have 
\begin{eqnarray}
    h(a,b)
    &\leq&\max\{h(0,0),h(0,1),h(1,0),h(1,1)\}\\
    &=&\max\{0,\frac{1}{2}(\ln\pi-\frac{3}{2}),\ln\pi-\frac{3}{2}\}\\
    &\leq&0.
\end{eqnarray}
Then combining with $\frac{\partial g(a,b,k)}{\partial k}<0$, for any $a,b\in\N$ and $k\in[1,2]$, we have 
\begin{eqnarray}
g(a,b,k)\leq g(a,b,1)=h(a,b)\leq0,
\end{eqnarray}
which implies $\frac{\partial f(a,b,k)}{\partial k}\leq0$. 

Thus, we conclude for any $(a,b)\in\N^2$ and $k\in[1,2]$, we have 
\begin{eqnarray}
f(a,b,k)\leq f(a,b,1)=0,
\end{eqnarray}
which proved the Eq. \eqref{even} for any $(a,b)\in\N^2$ and $k\in[1,2]$.

\section{Fock basis properties for states satisfying Condition \ref{cond1}}\label{appdxC}
\renewcommand{\thesubsection}{\arabic{subsection}}
\def\theequation{C\arabic{equation}}
\setcounter{equation}{0}
\renewcommand{\thefigure}{C\arabic{figure}}    
\setcounter{figure}{0}

In this Appendix, we show that Condition \ref{cond1} does not imply that the state $\hat{\rho}$ is Fock-diagonal but it implies $\hat{\rho}_{n,m}=0$ if $|n-m|$ is odd. 

Let $W(q,p)$ be a Wigner function satisfies Condition \ref{cond1} and $\chi_W(\eta)$, $\eta \in \mathbb{C}$, be its corresponding Wigner characteristic function. Without loss of generality, we can assume $n-m=l\geq0$ and get 
\begin{eqnarray}
    \langle n|\hat{\rho}|m\rangle
    &=&\frac{1}{\pi}\int\chi_W(\eta)\langle n|\hat{D}^{\dag}(\eta)|m\rangle d^2\eta\\
    &=&\frac{1}{\pi}\int\chi_W(\eta)e^{-\frac{|\eta|^2}{2}}\langle n|e^{-\eta\hat{a}^{\dag}}e^{\eta^*\hat{a}}|m\rangle d^2\eta\\
    &=&\frac{1}{\pi}\int\chi_W(\eta)e^{\frac{|\eta|^2}{2}}\left(\sum_{s=0}^{\infty}\frac{(-\eta^*)^s}{s!}\hat{a}^s|n\rangle\right)^{\dag}\left(\sum_{t=0}^{\infty}\frac{(\eta^*)^t}{t!}\hat{a}^t|m\rangle\right)d^2\eta\\
    &=&\frac{1}{\pi}\int\chi_W(\eta)e^{-\frac{|\eta|^2}{2}}\left(\sum_{s=0}^{n}\frac{(-\eta^*)^s}{s!}\sqrt{\frac{n!}{(n-s)!}}|n-s\rangle\right)^{\dag}\left(\sum_{t=0}^{m}\frac{(\eta^*)^t}{t!}\sqrt{\frac{m!}{(m-t)!}}|m-t\rangle\right)d^2\eta\\
    &=&\frac{1}{\pi}\int\chi_W(\eta)e^{-\frac{|\eta|^2}{2}}\left( \sum_{t=0}^{m}\frac{(-\eta)^{t+l}(\eta^*)^t}{(t+l)!t!}\frac{\sqrt{(m+l)!m!}}{(m-t)!} \right)d^2\eta\\
    &=&\frac{1}{\pi}\int\chi_W(\eta)e^{-\frac{|\eta|^2}{2}} (-\eta)^l \left( \sum_{t=0}^{m}\frac{(-1)^{t}|\eta|^{2t}}{(t+l)!t!}\frac{\sqrt{(m+l)!m!}}{(m-t)!} \right)d^2\eta.
\end{eqnarray}

The case for $n-m<0$ can be easily handled by switching $m$ to $n$ and $t$ to $s$. Since the (inverse) Fourier transform conserves the parity of the function, we have that $\chi_W(\eta)$ is an even function of $\eta$. Together with the fact that $\left( \sum_{t=0}^{m}\frac{(-1)^{t}|\eta|^{2t}}{(t+l)!t!}\frac{\sqrt{(m+l)!m!}}{(m-t)!} \right)$ is also an even function of $\eta$, the parity of the integrand only depends on $l$. Therefore, when $l$ is odd, which in turn means the integrand is odd in $\eta$, we have $\langle n|\hat{\rho}|m\rangle=0$. 

When $k$ is even, $\langle n|\hat{\rho}|m\rangle$ can be non-zero. The example below shows this. Consider a more general form of the Wigner function, $p_0|0\rangle\langle0|+p_1|1\rangle\langle1|+p_2|2\rangle\langle2|+c|0\rangle\langle2|+c^*|2\rangle\langle0|$, we have 
\begin{eqnarray}
    W(q,p)
    &=&W_0\left[(1-2p_1)+(2p_1-4p_2+2\sqrt{2}c_1)q^2+4p_2q^2p^2\right.\\\nonumber
    &&+\left.(2p_1-4p_2-2\sqrt{2}c_1)p^2+2p_2q^4+2p_2p^4-4\sqrt{2}c_2qp\right]. 
\end{eqnarray}
Therefore, all states with 
\begin{eqnarray}
p_1\leq\frac{1}{2},\quad p_1-2p_2-\sqrt{2}c_1\geq0,\quad c_2\leq0
\end{eqnarray}
satisfies Condition \ref{cond1}. So, when $p_0=\frac{1}{3},p_1=\frac{1}{2},p_2=\frac{1}{6},c=\frac{\sqrt{2}}{16}-i$, the state is not Fock-diagonal but still satisfies Condition \ref{cond1}.

\twocolumngrid
\bibliography{refs.bib}

\end{document}